\documentclass[aps,twocolumn,amssymb,superscriptaddress]{revtex4-1} 
\usepackage[final]{changes}
\usepackage[colorlinks=true,citecolor=blue,linkcolor=blue,urlcolor=blue]{hyperref}
\usepackage{amsmath}
\usepackage{amssymb}
\usepackage{bbm}
\usepackage{xcolor}
\usepackage{graphicx}
\usepackage{epstopdf}
\usepackage{dcolumn}
\usepackage{bm}
\usepackage{hyperref}
\graphicspath{{Latexpicture/}}

\begin{document}

\title{Deterministic single-photon source in the ultrastrong coupling regime}

\author{Jie Peng}
\email{jpeng@xtu.edu.cn}
\affiliation{Hunan Key Laboratory for Micro-Nano Energy Materials and Devices\\ and School of
Physics and Optoelectronics, Xiangtan University, Hunan 411105, China}

\author{Jianing Tang}
\affiliation{Hunan Key Laboratory for Micro-Nano Energy Materials and Devices\\ and School of
Physics and Optoelectronics, Xiangtan University, Hunan 411105, China}

\author{Pinghua Tang}
\affiliation{Hunan Key Laboratory for Micro-Nano Energy Materials and Devices\\ and School of
Physics and Optoelectronics, Xiangtan University, Hunan 411105, China}

\author{Zhongzhou Ren}
\affiliation{School of Physics Science and Engineering, Tongji University, Shanghai 200092, China}

\author{Junlong Tian}
\affiliation{College of Big Data and Information Engineering, Guizhou University, Guiyang 550025, China}

\author{Nancy Barraza}
\affiliation{International Center of Quantum Artificial Intelligence for Science and Technology (QuArtist) and Department of Physics,
Shanghai University, 200444 Shanghai, China}

\author{Gabriel Alvarado Barrios}
\affiliation{Kipu Quantum, Greifswalderstrasse 226, 10405 Berlin, Germany}

\author{Lucas Lamata}
\affiliation{Departamento de F\'isica At\'omica, Molecular y Nuclear, Universidad de Sevilla, 41080 Sevilla, Spain}
\affiliation{Instituto Carlos I de F\'isica Te\'orica y Computacional, 18071 Granada, Spain}

\author{Enrique Solano}
\email{enr.solano@gmail.com}
\affiliation{International Center of Quantum Artificial Intelligence for Science and Technology (QuArtist) and Department of Physics,
Shanghai University, 200444 Shanghai, China}
\affiliation{Kipu Quantum, Greifswalderstrasse 226, 10405 Berlin, Germany}
\affiliation{IKERBASQUE, Basque Foundation for Science, Plaza Euskadi 5, 48009 Bilbao, Spain}

\author{F. Albarr\'an-Arriagada}
\affiliation{Departamento de F\'isica, Universidad de Santiago de Chile (USACH), Avenida V\'ictor Jara 3493, 9170124, Estaci\'on Central, Chile}
\affiliation{Center for the Development of Nanoscience and Nanotechnology, 9170124, Estaci\'on Central, Chile}
\begin{abstract}
Deterministic single-photon sources are important and ubiquitous in quantum information protocols. However, to the best of our knowledge, none of them work in the ultrastrong light-matter coupling regime, and each excitation process can only emit one photon. We propose a deterministic single-photon source in circuit QED which can work in the ultrastrong coupling regime. Here, two qubits are excited simultaneously in one process and two deterministic single photons can be sequentially emitted with an arbitrary time separation. This happens through two consecutive adiabatic transfers along the one-photon solutions of the two-qubit Rabi and Jaynes-Cummings model, which has constant eigenenergy in the whole coupling regime. Unlike the stimulated Raman adiabatic passage, the system goes back to the initial state of another period automatically after photon emission. Our scheme can approach unity single-photon efficiency, indistinguishability, and purity simultaneously. With the assistance of the Stark shift, a deterministic single photon can be generated within a time proportional to the inverse of the resonator frequency. 
\end{abstract}
\maketitle

\emph{Introduction.--}
Single photon sources are fundamental building blocks in quantum information, with applications ranging from quantum computation \cite{klm,kok} to quantum communication \cite{cirac,pan1,pan2} and sensing \cite{degen}. The different technologies for single-photon sources can be classified into two families \cite{thomas}. The first one considers a nonlinear material process such as spontaneous parametric
downconversion \cite{lou,burn}, where a pump laser illuminates a material with a $\chi(2)$
optical nonlinearity, creating two photons, one of which can be used to herald the creation of the other. However, this process is probabilistic and inefficient \cite{bon,thomas}. The second approach is based on single quantum emitters, which deterministically emit one photon at a time. This has been demonstrated in atoms \cite{Darquie}, molecules \cite{Christian}, ions \cite{Maunz}, color centers in diamonds \cite{Kurtsiefer}, and quantum dots \cite{Charles}. However, the spontaneous emission in all directions makes photon collection difficult. One way to solve this problem is to couple the emitters to a cavity to enhance the radiation into the cavity mode, so that the photon can be emitted from the cavity through a certain direction \cite{ion}. Meanwhile, the emission rate can be increased by Purcell effect \cite{circuit}, and the pure dephasing is reduced due to the coupling to a fixed cavity mode, making the photons more indistinguishable \cite{tunable}. Using a quantum dot coupled to cavities, near-unity indistinguishability and purity are realized simultaneously, with an extraction efficiency of $66\%$ \cite{dx}, a polarized single-photon efficiency of $60\%$ \cite{wang} and an overall efficiency of $57\%$ \cite{tomm}, respectively. Besides high efficiency \cite{tunable}, purity \cite{Bozyigit} and indistinguishability \cite{lang}, a microwave single photon source realized in circuit QED can also be tunable \cite{tunable,peng}. 

The Purcell effect, which enhances the emission rate, is proportional to the qubit-cavity coupling strength \cite{Kaer}. When we enter the strong coupling regime, the swap between the qubit excited state and the single photon state in the cavity will be accelerated. Since now the ultrastrong \cite{ultra} and even deep-strong coupling \cite{deep} have been realized in circuit QED, it is natural to consider implementing a single photon source using these stronger coupling strengths, which most likely will accelerate the photon generation speed. However,  to the best of our knowledge, this has not been realized yet. Because the ultrastrong coupling will invoke counter-rotating terms, which excite the qubit and photon simultaneously, all photon number states become connected, and the system, described by the quantum Rabi model \cite{braak,chen}, normally involve an infinite number of photons \cite{zhong,xie,yu}. 

Recently, we have found special dark states of the two-qubit quantum Rabi model \cite{pj1,pj2,pj3}, which contain at most one photon, and have constant eigenenergy in the whole coupling regime. Here, we propose to implement a deterministic single photon source that takes advantage of the ultrastrong coupling using these dark states. Our scheme begins with qubits and the cavity ground state, and then the qubits are excited simultaneously. Next, the system undergoes an adiabatic evolution along the aforementioned dark state, and end up with a product state of a single photon state and a two-qubit singlet state.  Because the reaching of ultrastrong coupling and peculiarity
of the dark state, the target state can be generated with fidelity $99.2\%$ in a time of $68 \omega^{-1}$, where $\omega$ is the cavity frequency. Furthermore, with the addition of Stark shift terms \cite{Grimsmo1,Grimsmo2,cong}, the minimum energy gap between the dark state and its closest eigenstates can be increased to $\approx 0.54\omega$, which accelerates the adiabatic evolution to $12\omega^{-1}$ with fidelity reaching $99\%$, showing a sign of ultrafast state generation \cite{ultrafast}. Moreover, after the photon is emitted, the system will be in an eigenstate of the two-qubit JC model with the excitation number $C=1$, which has also constant eigenenergy in the whole
coupling regime. So through another adiabatic process, a single photon can be generated and emitted through dissipation, leaving the system in its ground state, which is also the initial state of a period. Unlike the STIRAP, no recycling of the qubit between photon generations is
required, which will increase the repetition rate \cite{Wilk}. Here one excitation process emits two single photons with efficiencies over 99\% with an arbitrary time separation, and their purity and indistiguishability both approach unity. Finally, we propose an experimental realization in circuit QED.

\begin{figure}[htbp]
\centering 
\resizebox{1\columnwidth}{!}{
  \includegraphics{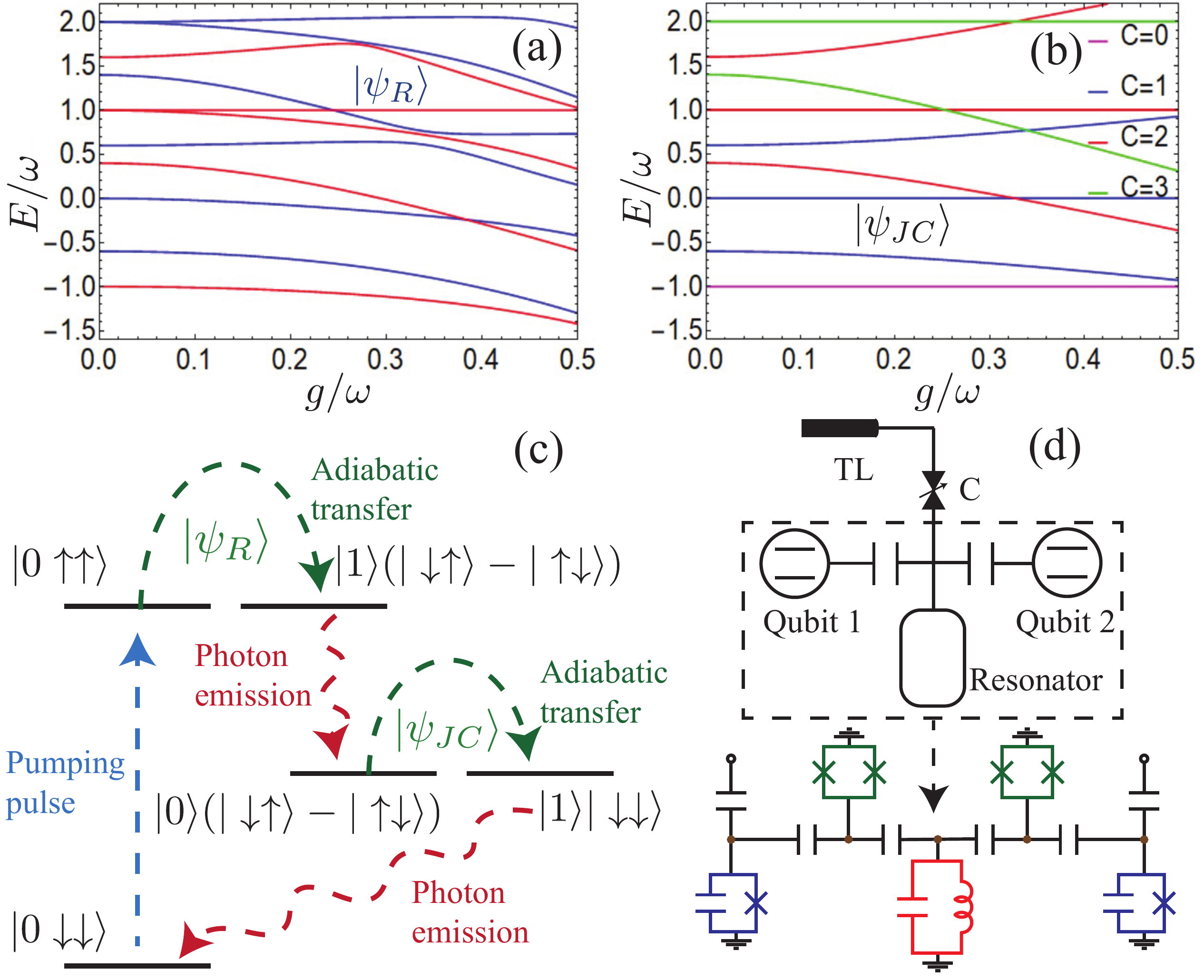}
}
\caption[1]{(a) The numerical spectrum of two-qubit quantum Rabi model with $\Delta_1=0.8\omega$, $\Delta_2=0.2\omega$, $g=g_1=g_2$. The red lines have even parity while the blue lines have odd parity. (b) The spectrum of two-qubit JC model with the same parameters. (c) Relevant energy levels and transitions of our  scheme. (d) Setup for the scheme: Two superconducting qubits are coupled to one resonator, whose photon emission rate into the TL is controlled by a variable coupler C. The lower part is a superconducting circuit design for the two-qubit Rabi and JC models with tunable couplings and qubit frequencies \cite{sl}. } 
\label{fig.1}  
\end{figure}

\emph{Scheme and circuit QED implementation.--}
Our scheme of the deterministic single-photon source is based on the one-photon solutions to the two-qubit Rabi and JC models ($\hbar=1$)
\begin{eqnarray}\label{eq2}
H_{Rabi}&=&\omega a^{\dag }a + g_{1}{\sigma _{1x}}(a + {a^\dag }) + g_{2}{\sigma _{2x}}(a + a^{\dag })\nonumber\\&& + \Delta _{1}\sigma _{1z} + \Delta _{2}{\sigma _{2z}},\\
H_{JC}&=&\omega a^\dag a+g_{1}(\sigma_{1}^\dag a+\sigma_{1} a^\dag)+g_{2}(\sigma_{2}^{\dag} a+\sigma_{2} a^\dag)\nonumber\\&& +\Delta_{1}\sigma_{1z}+\Delta_{2}\sigma_{2z} ,
\end{eqnarray}
where $a^{\dagger}$ and $a$ are the photon creation and annihilation operators with frequency $\omega$, respectively. Also, $\sigma_{j \alpha} (\alpha=x,y,z)$ are the Pauli matrices corresponding to the $j$-th qubit. $2\Delta_j$ is the energy level
splitting of the $j$-th qubit, and $g_{j}$ is the qubit-photon coupling parameter between the resonator and $j$-th qubit.
Normally, there is no solution with finite photon numbers to $H_{Rabi}$ since all the photon number states are connected. However, we have found special dark states with at most one-photon~\cite{pj1}
\begin{eqnarray}\label{psir}
\vert\psi_{R}\rangle=\frac{1}{{\cal N}}[(\Delta_1 - \Delta_2)\vert 0\uparrow\uparrow\rangle+g\vert 1\rangle(\downarrow\uparrow\rangle-\uparrow\downarrow\rangle)] ,
\end{eqnarray}
where $\Delta_1+\Delta_2=\omega$ and $g_1=g_2=g$, corresponding to the horizontal line $E=\omega$ in Fig. \ref{fig.1}(a). Surprisingly, horizontal lines $E/\omega=N~(N=0,1,2,\ldots)$  emerge in the energy spectrum of the two-qubit JC model under the same conditions, as shown in Fig. \ref{fig.1}(b). 

The corresponding eigenstates read \cite{pj4}
\begin{eqnarray}\label{psijc}
\vert\psi_{JC}\rangle=\frac{1}{{\cal N^{\prime}}}[(\Delta_1 - \Delta_2)\vert 1\downarrow\downarrow\rangle+g(\vert 0\downarrow\uparrow\rangle-\vert 0\uparrow\downarrow\rangle)]
\end{eqnarray} 
for $N=0$ \cite{pj4}. We will show that $\vert\psi_{R}\rangle$ and $\vert\psi_{JC}\rangle$ can be used to produce a special single-photon source through two consecutive adiabatic transfers.  

Our scheme is depicted in Fig. \ref{fig.1} (c) and (d). Two qubits are coupled to one resonator, which is connected to a transmission line (TL) through a variable coupler C, so that its dissipation rate $\kappa_c$ is tunable. The qubits and the resonator are cooled down to the ground state $\vert 0\downarrow\downarrow\rangle$ initially. Then pumping pulses, 
\begin{equation}
H=\frac{\Omega}{2}(\sigma_{1}^\dag{e^{ - i\omega_{q1} t }} +\sigma_{1}{e^{  i\omega_{q1} t }}+\sigma_{2}^\dag{e^{ - i\omega_{q2} t }} +\sigma_{2}{e^{  i\omega_{q2} t }}) ,
\end{equation}
are applied to excite the qubits, where $\omega_{q1}$, $\omega_{q2}$ are the frequencies of the two qubits, respectively. Initially, we set $\omega_{q1}=2\Delta_1\approx 2\omega$, $\omega_{q2}=2\Delta_2\approx 0$, so the crosstalk can be neglected and the coupling $g=g_1=g_2$ is set to zero. After excitation, $\vert 0\uparrow\uparrow\rangle$ just corresponds to $\vert \psi_R\rangle$ of Eq. \eqref{psir} since $g=0$ and $\Delta_1\neq\Delta_2$. Next, we increase $g$ to a nonzero value and decrease $\Delta_1-\Delta_2$ to $0$, so that the state $\vert 0\uparrow\uparrow\rangle$ evolves adiabatically to $\vert 1\downarrow\uparrow\rangle-\vert 1\uparrow\downarrow\rangle$ through $\vert \psi_R\rangle$. Then, we activate C to increase the dissipation rate into the TL to a very large value, which can be achieved in a few
\begin{figure}[htbp]
\centering 
\resizebox{1\columnwidth}{!}{
  \includegraphics{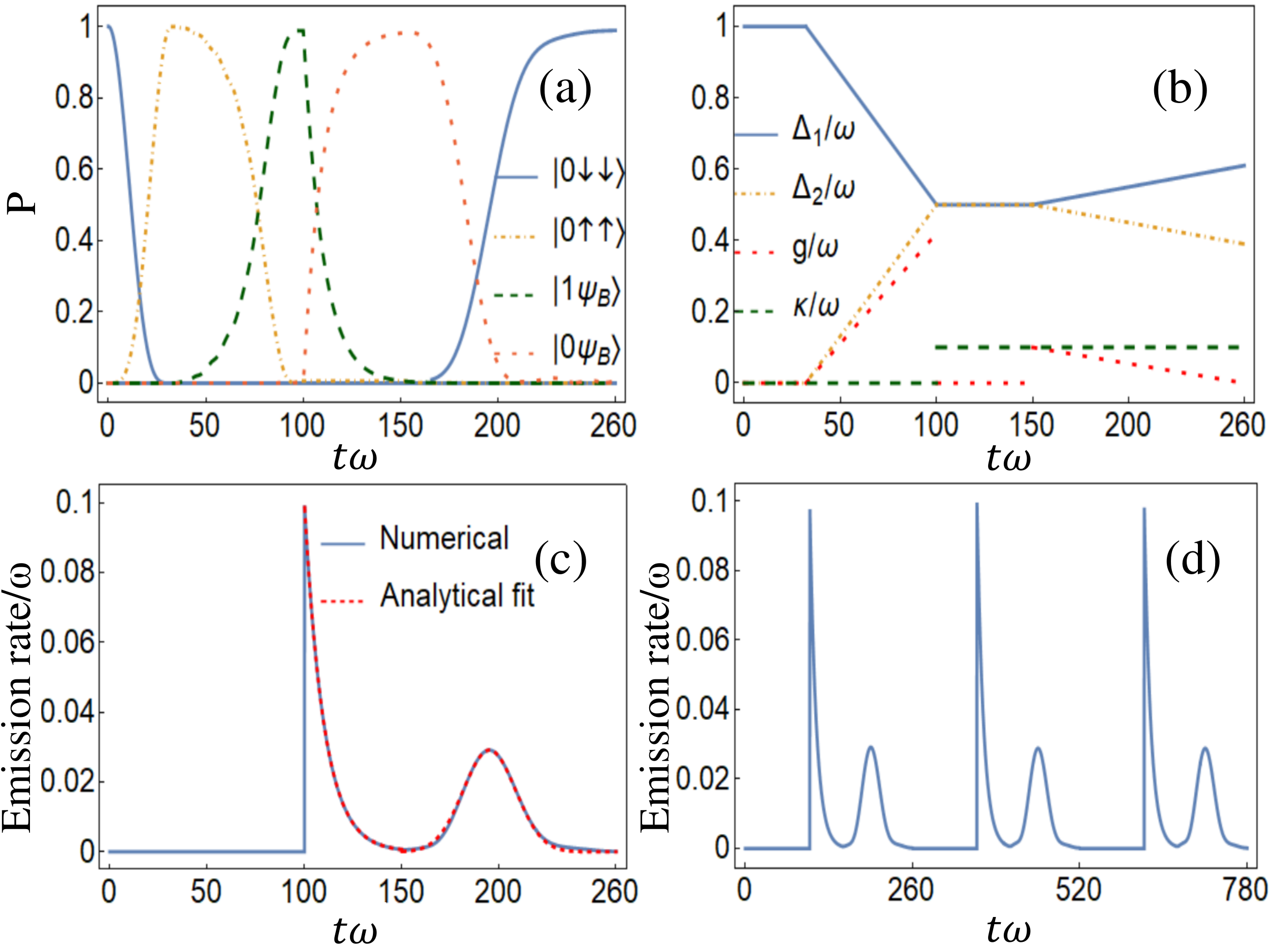}
}
\caption[1]{(a) Shows the evolution of system states during one period, where $\vert \psi_B\rangle=\vert \downarrow\uparrow-\uparrow\downarrow\rangle$. (b) The corresponding evolution of the  parameters. (c) The waveforms of the emitted photons. The solid line corresponds to the numerical simulation, fitted by an exponential decaying function $\exp(-\kappa t)$ and a Gaussian function, represented by the dashed line. (d) A sequence of single photons are generated when we repeat the above processes.} 
\label{fig.2}  
\end{figure}  nanoseconds \cite{catch}. The single photon can be emitted into the TL and the system state becomes $\vert 0\downarrow\uparrow\rangle-\vert 0\uparrow\downarrow\rangle$, corresponding to the one-photon solution $\vert \psi_{JC}\rangle$ of Eq.~\eqref{psijc} for the two-qubit JC model. After an arbitrary time separation, we begin another adiabatic evolution along $\vert \psi_{JC}\rangle$ by decreasing $g$ to zero and increasing $\Delta_1-\Delta_2$ to a nonzero value, which generates $\vert 1\downarrow\downarrow\rangle$, and the single photon can be emitted during this process. After that, the system returns to the initial state $\vert 0\downarrow\downarrow\rangle$ which can be utilized for the next period automatically. Furthermore, we propose a superconducting circuit design for the two-qubit Rabi and JC models with tunable couplings and qubit frequencies, shown in the lower part of Fig. \ref{fig.1} (d), with a detailed demonstration in \cite{sl}.

Usually, one excitation process can only emit one photon for deterministic single-photon sources, and the system has to be repumped afterwards, so there is always an unavoidable time separation between the photons emitted. However, with our protocol one pumping process can emit two single photons, since the system reaches the ``excited state'' of the second emission process automatically after the first photon is emitted, which also provides an arbitrarily controllable time separation of the photons. 

We simulate the process described above using the Lindblad master equation, with the resonator intrinsic dissipation rate $\kappa_{in}=10^{-4}\omega$, the qubit damping rate $\gamma=10^{-5}\omega$ and dephasing rate $\gamma_{\phi}=2\times10^{-5}\omega$ \cite{sl}. The evolution of different states in a period is shown in Fig. \ref{fig.2} (a), with corresponding parameters depicted in Fig. \ref{fig.2} (b). The absolute values of the detunnings of both qubits with respect to the resonator are always the same during the evolution, which implies that the magnetic flux used to tune the qubit frequencies can also be used to tune the coupling strength, maintaining $g_1=g_2$ \cite{sl}. This will make our implementation easier to realize.  Note that $g$ can be arbitrary values during $100\omega^{-1}-150\omega^{-1}$, because the qubit singlet state is decoupled from the cavity when emitting the first photon. Here, we choose $g=0$ to have a better single photon purity. Although the fast adiabatic evolution along $\vert \psi_R\rangle$ may seem difficult according to the adiabatic theorem, which states that the system will evolve along $\vert\psi_{R}\rangle$ if
\begin{equation}\label{adiabatic}
\left|\frac{\langle E_n(t)|\dot{H}_{R}|\psi_{R}(t)\rangle}{(E_n-E_{R})^2}\right|\ll1,
\end{equation}
where there is an energy level $\vert E_n\rangle$ very close to $\vert\psi_{R}\rangle$, as shown in Fig. \ref{fig.1} (a). However, $\vert 1 (\downarrow\uparrow-\uparrow\downarrow)\rangle$ can be fast generated from $\vert \uparrow\uparrow\rangle$ in $68 \omega^{-1}$ with a fidelity of $99.2\%$ through the adiabatic evolution along $\vert \psi_R\rangle$. This is mainly because the peculiarity of the dark state and reaching of ultrastrong coupling regime \cite{sl}. The photon emission rate into the TL is shown in Fig. \ref{fig.2} (c), where the time separation between two emitted photons is zero. The first photon has a typical single-sided exponential waveform decaying with the lifetime of the resonator, while the second photon has a Gaussian shape which is optimal for tolerance to mode mismatch \cite{Sonia}. We use an exponential function $\exp(-\kappa t)$ and a Gaussian function to fit the simulated data and find good consistency, as can be seen in Fig. \ref{fig.2} (c). The system is reset to $\vert 0\downarrow\downarrow\rangle$ automatically after the second photon emission, so we can repeat the above process to obtain a sequence of single photons, as shown in Fig. \ref{fig.2} (d). 

\emph{Figures of merit--}
There are commonly three important figures of merit for a single photon source \cite{Reimer}: efficiency, purity and indistinguishability. First, the generation and collection efficiency are defined as the photon generated and collected for one excitation pulse, respectively, which is equal to one for a perfect single photon source. Here, two single photons are generated in one excitation process, with emission probabilities (efficiencies), $\kappa_{c}\int a^{\dag}(t) a(t) {\rm d}t$, both larger than $99\%$. The generation and collection efficiencies both approach unity, since the qubits are strongly coupled to the resonator, and their interaction is much faster than the decoherence rate of the qubit. The single photon is generated in the resonator and almost fully collected by the TL through the variable coupler because $\kappa_c$ can be tuned to be $1000$ times the intrinsic decay rate in a few nanoseconds \cite{catch}. 
 
\begin{figure}[htbp]
\centering 
\resizebox{1\columnwidth}{!}{
  \includegraphics{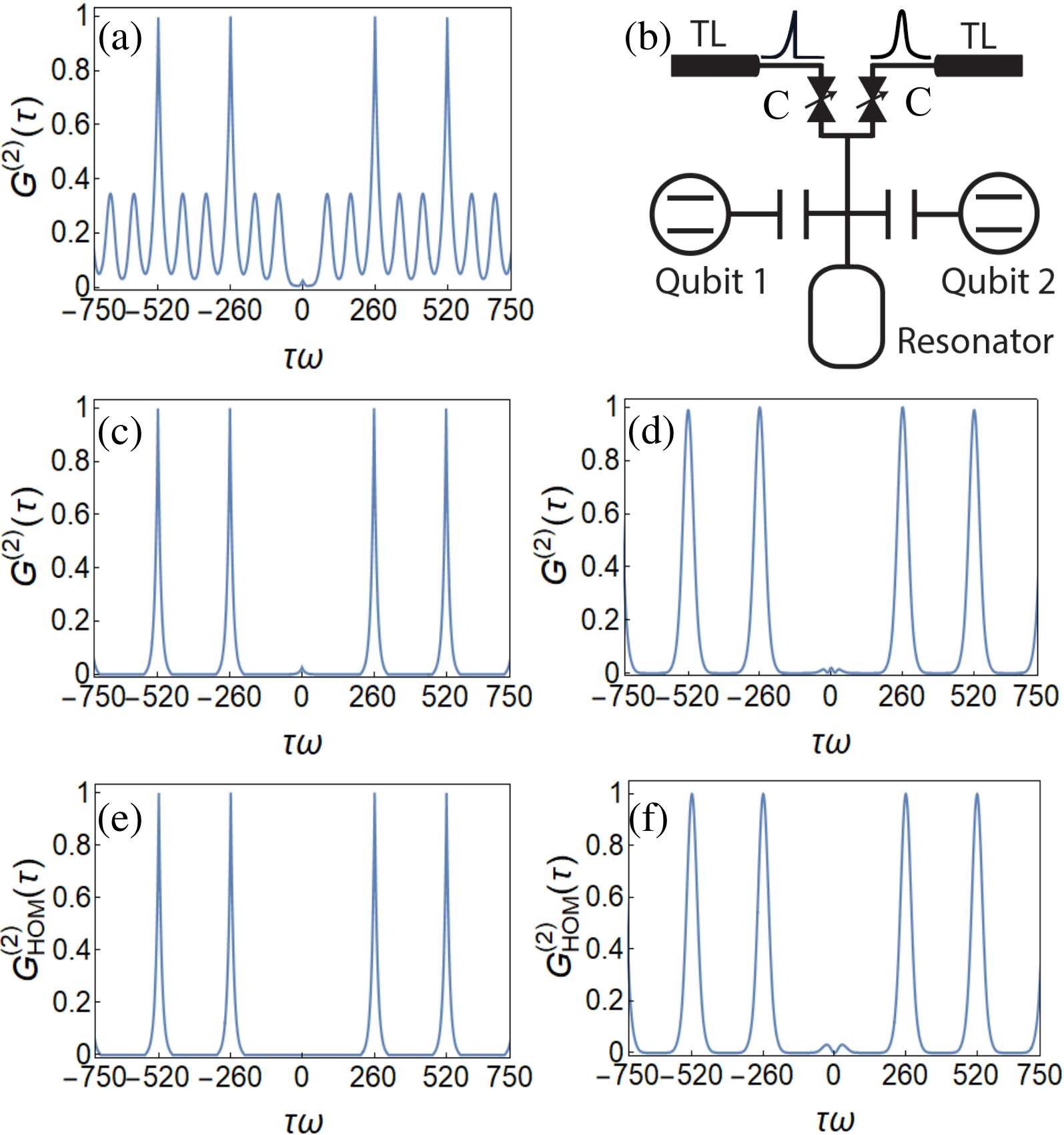}
}
\caption[1]{(a)The simulated second-order correlation function in HBT experiment for the single-photon source proposed above, which characterizes its purity. (b) Our scheme to collect two single photons separately. (c) The second-order correlation function in HBT experiment for the first single photon source. (d) The second-order correlation function in HBT experiment for the second single photon source. (e) The second-order correlation function in HOM experiment for the first single photon source, which characterizes its indistinguishability. (f) The second-order correlation function in HOM experiment for the second single photon source. All these simulated correlation functions are normalized to the highest peak.} 
\label{fig.3}  
\end{figure}
The second requirement for a single photon source is high purity, which means exactly one photon
is emitted at a time. It can be characterized by the second-order correlation functions in the Hanbury-Brown-Twiss (HBT) experiment
\begin{eqnarray}
G^{(2)}(t, \tau ) &=&\langle {a^\dag }(t){a^\dag }(t + \tau )a(t + \tau )a(t)\rangle,\nonumber\\
G^{(2)}( \tau )&=&\int G^{(2)}(t, \tau )dt,
\end{eqnarray}
where the former represents the probability of detecting a photon at time $t$ and another one at time $t+\tau$, and the later is its integration over $t$, as shown in Fig. \ref{fig.3} (a). The data has been normalized to the highest peak. There is a strongly suppressed $G^{(2)}( 0 )\approx 0.025$, giving a clear evidence of the single-photon emission. $G^{(2)}( \tau )$ has the same period as the single-photon source, and two different kinds of peaks other than $\tau=0$, because the waveforms of photons in a period are not identical. The first peak arises at $\tau\approx 88\omega^{-1}$, corresponding to the coincidence detection of the first and the second photons, and the second peak arises at $\tau\approx 173\omega^{-1}$, which has the same height as the first peak caused by the coincidence detection of the second and the third photons. The highest peak arises clearly at $\tau\approx 260\omega^{-1}=T_{period}$ because of the periodicity.
 
The third requirement is high indistinguishability, which means the photons must be identical in all degrees of freedom, to ensure the successful implementation of the two-photon gate through interference. The Hong, Ou and Mandel (HOM) experiment \cite{hom} is used to characterize this notion, where two photons are interfered on a 50/50 beam splitter. If they are completely indistinguishable, they will coalesce and exit through the same beam splitter output. Here two single photons are emitted in one period with different temporal waveforms which may be made identical with the photon waveform reshaping technology \cite{Kie} to implement a two-photon CNOT gate. Since the time separation between these two single photons is arbitrary, we have much freedom to choose the delay time between the control and target photons. 

\begin{figure}[htbp]
\centering 
\resizebox{1\columnwidth}{!}{
  \includegraphics{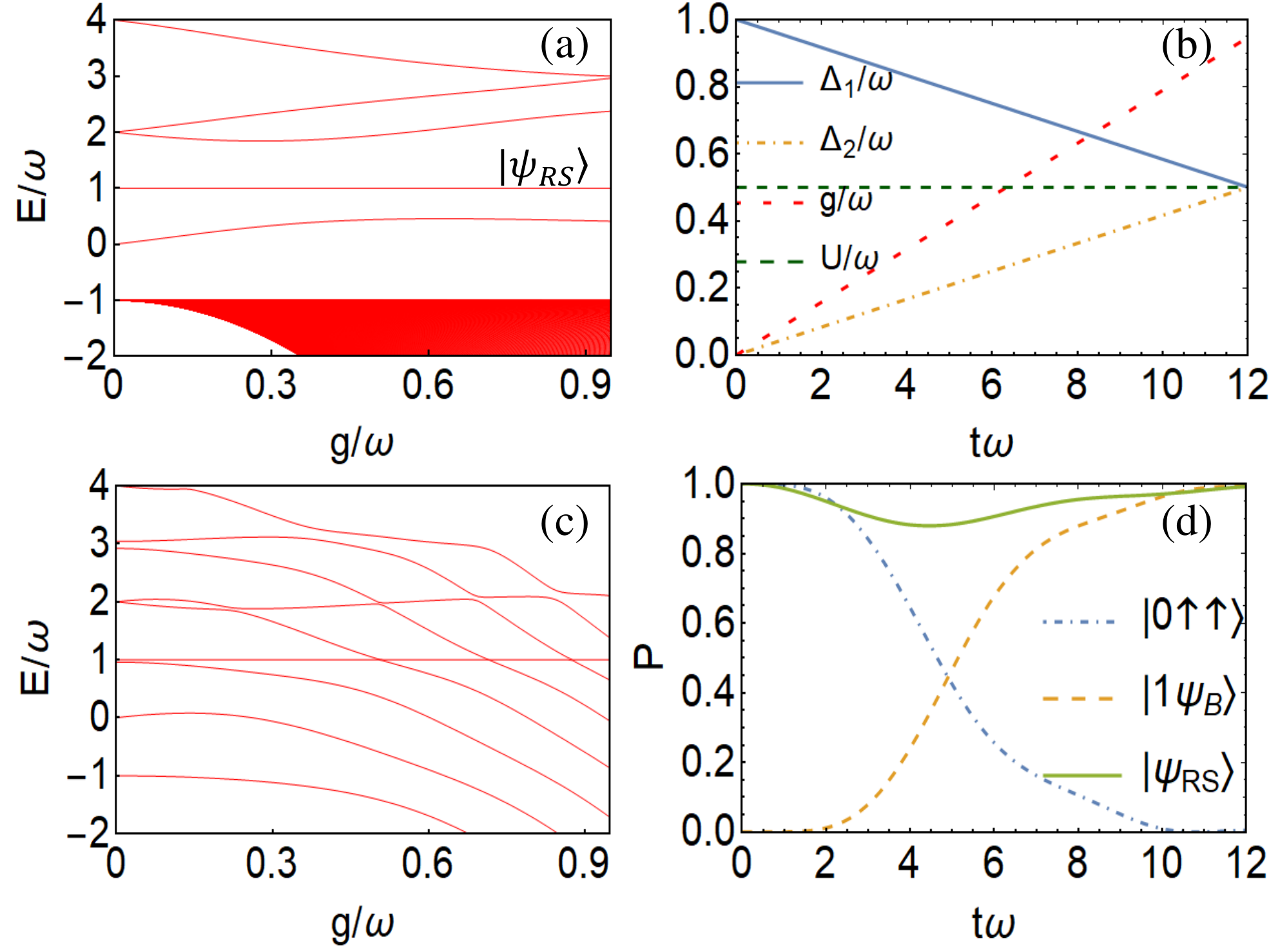}
}
\caption[1]{(a) The numerical spectrum of $H_{RS}$ with even parity when parameters evolve according to (b). (b) The time-dependent parameters during the adiabatic evolution from $\vert 0\uparrow\uparrow\rangle$ to $\vert 1\psi_B\rangle$ along $\vert\psi_{RS}\rangle$, where $U=U_1=U_2=0.5\omega$. (c) The numerical spectrum of $H_{RS}$ of even parity with the same parameters as (a) except for $U_1=U_2=0.01\omega$. (d) Ultrafast generation of $\vert 1\psi_B\rangle$ through adiabatic evolution along $\vert\psi_{RS}\rangle$, with parameters evolving according to (b).} 
\label{fig.4}  
\end{figure}

To make better use of this single photon source, we can add another output channel to the resonator with a variable coupler to collect these two single photons separately, as shown in Fig. \ref{fig.3} (b) , so that the photons in each channel are indistinguishable, and with only one excitation process in a period, we efficiently generate two sequences of indistinguishable single photons. The first one has a typical single-sided exponential waveform. Its simulated $G^{(2)}( \tau )$ is shown in Fig. \ref{fig.3} (c), with $G^{(2)}(0)\approx 0.026$.  The second one has a Gaussian shape, whose $G^{(2)}( \tau )$ is shown in Fig. \ref{fig.3} (d), with $G^{(2)}(0)\approx 0.02$. They can be applied in different situations. We also simulated the normalized HOM experiment counts $G^{(2)}_{HOM}(\tau)$ of photon detection in different outputs at time delay $\tau$. There is a clear vanishing second order correlation function $G^{(2)}_{HOM}(0)$ for the first photon shown in Fig. \ref{fig.3} (e) because of the indistinguishability, while typical small peaks around $\tau=0$ arise for the second photon in Fig. \ref{fig.4} (f), because the decoherence of the qubits cause random photon frequency differences. However, the first photon has already been generated before emission, and its spatio-temporal coherence is governed by resonator decay, showing no significant additional dephasing \cite{lang}.
The indistinguishability is defined by dividing the area of the peaks around $\tau=0$ by that of the uncorrelated peak around $\tau=T$, and subtracting this number from unity. It can be calculated as \cite{Kaer}
\begin{equation}
I=\frac{ 
\int_0^\infty {\rm d} t \int_0^\infty  {\rm d}\tau |\langle a^{\dag} (t+\tau) a(t)\rangle|^2}{\int_0^\infty  {\rm d} t \int_0^\infty  {\rm d}\tau \langle a^\dag(t+\tau) a(t+\tau)\rangle\langle a(t)^{\dag} a(t)\rangle}.
\end{equation} 
Because of periodicity, we only need to consider the integration over one period \cite{quantum-dot}. The indistinguishability, $I$, reaches $99.99\%$ for the first photon and $96.1\%$ for the second photon.

Another way to generate indistinguishable photons is to turn off the variable coupler $C$ during the adiabatic evolution from $\vert 0\rangle(\vert \downarrow\uparrow\rangle-\vert \uparrow\downarrow)$ to $\vert 1 \downarrow\downarrow\rangle$, so the second photon will be generated inside the resonator and then emitted into the TL when $C$ is turned on, just like the first photon. These two single photons will clearly be indistinguishable, with $I\approx1$.

\emph{Ultrafast generation of the single photon with the assistance of the Stark shift.--}
Another requirement for the single photon source is high speed, meanwhile, the most prominent advantage of the ultratrong coupling is the ultrafast state generation, which can be realized here by adding Stark shift terms to the two-qubit Rabi model \cite{Eckle,yf,li1,li2}
\begin{eqnarray}\label{eq2}
H_{RS}&=&\omega a^{\dag }a + g_{1}{\sigma _{1x}}(a + {a^\dag }) + g_{2}{\sigma _{2x}}(a + a^{\dag })\nonumber\\&& + \Delta _{1}\sigma _{1z} + \Delta _{2}{\sigma _{2z}}+U_1 a^{\dag }a\sigma_{1z}+U_2 a^{\dag }a\sigma_{2z},
\end{eqnarray}
where $U_1$ and $U_2$ are couplings of the stark terms. The photon frequency is shifted by $U_1\sigma_{1z}+U_2 \sigma_{2z}$, so the stability of the system requires $U_1+U_2\leq \omega$. We find a similar special dark state
\begin{eqnarray}
\vert\psi_{RS}\rangle&=&\frac{1}{{\cal N}}[(\Delta_1 - \Delta_2+U_1-U_2)\vert 0\uparrow\uparrow\rangle\nonumber\\
&&+g(\vert 1\downarrow\uparrow\rangle-\vert 1\uparrow\downarrow\rangle)]
\end{eqnarray}
with constant energy $E=\omega$ in the whole coupling regime, under the same condition $\Delta_1+\Delta_2=\omega$ and $g=g_1=g_2$ as for $H_{Rabi}$ \cite{sl}. 
Interestingly, the energy gap between $\vert\psi_{RS}\rangle$ and its closest eigenstates is enlarged by the stark terms. E. g., $\vert\psi_{RS}\rangle=\vert0\uparrow\uparrow\rangle$ and $\vert 2\downarrow\downarrow\rangle$ are degenerate for $H_{Rabi}$ at $g=0$, but the stark terms will reduce the energy of the latter by $U_1+U_2$, so these two energy levels are separated in the spectrum.  This phenomena is quite significant and interesting when $U_1+U_2=\omega$, where all $\vert n \downarrow\downarrow\rangle$ become degenerate ground states at $g=0$ with $E=-\omega$, as shown in Fig. \ref{fig.4} (a).
These infinite many energy levels lie below $E=-\omega$ in the spectrum, which should have appeared in the upper part if  $U_1+U_2<\omega$, as can be seen from Fig. \ref{fig.4} (c). All these states become quite far
 from $\vert\psi_{RS}\rangle$, and the minimum gap between  $\vert\psi_{RS}\rangle$ and other states $\vert E_n\rangle$ is about $0.54~\omega$, as can be seen in Fig. \ref{fig.4} (a). Therefore, according to the adiabatic theorem Eq. \eqref{adiabatic}, the adiabatic process should be much faster than for cases where the minimum gap is much smaller than $\omega$. With the parameters changing as Fig. \ref{fig.4} (b),  $\vert 1\psi_B\rangle$ can be generated from $\vert 0\uparrow\uparrow\rangle$ through an adiabatic evolution along $\vert \psi_{RS}\rangle$ with fidelity $99.2\%$, operating at a time proportional to the inverse of the resonator frequency $f$ ($12\omega^{-1}= \frac{6}{\pi}f^{-1}$), which is a sign of ultrafast quantum-state generation \cite{ultrafast}, as shown in Fig. \ref{fig.4} (d). Note that many other choices of $U$ also significantly accelerate the adiabatic evolution. E. g., $\vert 1\psi_B\rangle$ can be generated in $12\omega^{-1}$ with fidelity $99\%$ for $U_1=U_2=0.45\omega$. Following the same procedure detailed above, a faster single-photon source can be implemented. 

\emph{Conclusion.--}
We have proposed a scheme for a deterministic single-photon source that can work in the ultrastrong coupling regime. That is through two consecutive adiabatic transfers along the one-photon solutions of the two-qubit Rabi model and JC model, respectively. An important advantage of our scheme is that one pumping process can emit two deterministic single photons with an arbitrary time separation that is easily controlled. Furthermore, with our protocol, the system evolves naturally to the initial state of the next period after photon emission. We characterize our single photon source by calculating efficiency, purity, and indistinguishability, showing that all of them can reach near unity values simultaneously. Moreover, by introducing Stark shift terms, we can accelerate the speed of the single-photon generation to a degree proportional to the inverse of the resonator frequency. Our scheme paves the way for the application of ultrastrong coupling in fast computation and deterministic state generation.

\emph{Acknowledgements.--}
We acknowledge Guillermo Romero for helpful discussions. This work was supported by the Natural Science Foundation of Hunan Province, China (2022JJ30556), the National Natural Science Foundation of China (Grant No. 11874316), the National Basic Research Program of China (2015CB921103), the Program for Changjiang Scholars and Innovative Research Team in University (No. IRT13093), Junta de Andaluc\'{i}a (P20\_00617 and US-1380840), ANID Subvenci\'on a la Instalaci\'on en la Academia SA77210018, ANID Proyecto Basal AFB 180001.

\newpage
\newcommand{\red}[1]{\textcolor{red}{#1}}
\newcommand{\blue}[1]{\textcolor{blue}{#1}}
\newcommand{\green}[1]{\textcolor{green}{#1}}
\newcommand{\magenta}[1]{\textcolor{magenta}{#1}}
\newcommand{\cyan}[1]{\textcolor{cyan}{#1}}
\newcommand{\brown}[1]{\textcolor{brown}{#1}}
\newcommand{\ket}[1]{\vert #1 \rangle}
\newcommand{\bra}[1]{\langle #1 \vert}
\newcommand{\ketbra}[2]{\vert #1 \rangle \langle #2 \vert}
\newcommand{\braket}[2]{\langle #1 \vert #2 \rangle}
\newcommand{\eg}{{\it{e.g.}}}
\newcommand{\ie}{{\it{i.e.}}}
\begin{widetext}

\subsection*{}
{\bf \large Supplementary Material: Deterministic single photon source in the ultrastrong coupling regime}

\renewcommand{\thesection}{S\arabic{section}}
\renewcommand{\thesubsection}{\Alph{subsection}}
\renewcommand{\thesubsubsection}{\alph\arabic{subsubsection}}
\renewcommand{\theequation}{S\arabic{equation}}
\renewcommand{\thefigure}{S\arabic{figure}}
\renewcommand{\thetable}{S\arabic{table}}
\setcounter{equation}{0}
\setcounter{figure}{0}

~\\

This supplementary material contains four parts: (1) Demonstration of the circuit design for the implementation of the two-qubit quantum Rabi and Jaynes-Cummings (JC) model with independently tunable couplings and qubit frequencies. (2) The Lindblad master equation used for numerical simulation. (3) Peculiarities of the special dark state $\vert \psi\rangle$ for the valid and fast adiabatic evolution along it. (4) Method to obtain the one-photon special dark-sate solution to the Rabi-Stark model.

\section{Circuit design for the two-qubit quantum Rabi and JC models }
\begin{figure}[b]
\centering
\includegraphics[width=0.6\linewidth]{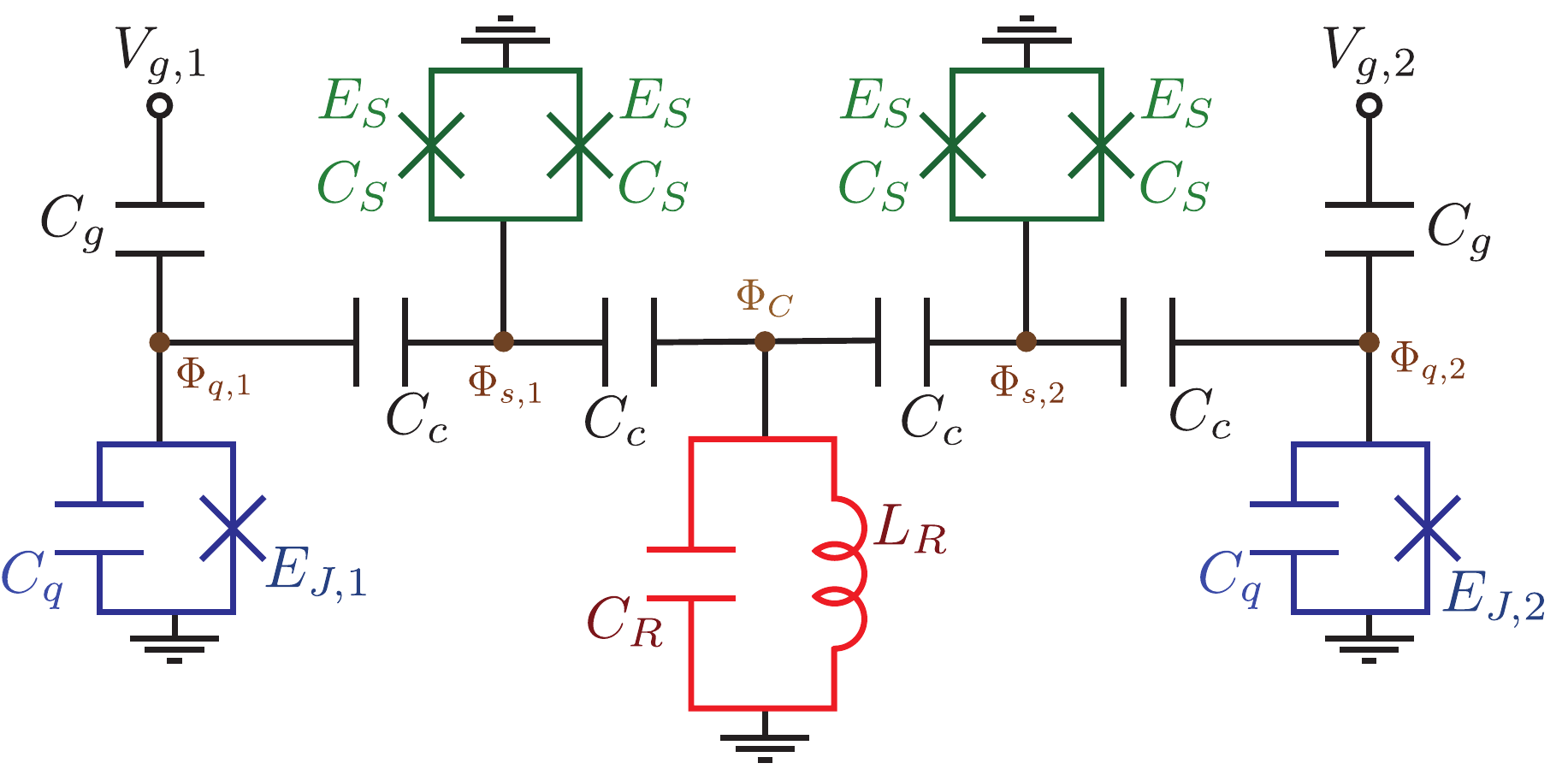}
\caption{Circuit diagram for the implementation of the two-qubit quantum Rabi model and JC models with tunable coupling and qubit frequency.}
\label{Fig01}
\end{figure}
The experimental implementation of our protocol for a single-photon source implies the implementation of the two-qubit quantum Rabi model with tunable coupling and tunable qubit frequency. One of the platforms that allow us to implement tunable coupling between qubits and resonators is superconducting circuits, where also the qubits have a tunable energy gap, and time-dependent coupling strength can be engeniered~\cite{Blais2020NatPhys, Devoret2013Science, Blais2021RevModPhys}. In particular, we propose the circuit design given in Fig.~\ref{Fig01} which is the single oscillator version of the circuit design presented in Ref.~\cite{Peng2021PhysRevLett}.

This circuit is described by the Lagrangian
\begin{eqnarray}\nonumber
\mathcal{L} = && \sum_{j=1}^2\bigg[\frac{C_{g}}{2}(\dot{\Phi}_{q,j}-V_{g,j})^2+\frac{C_{q}}{2}\dot{\Phi}_{q,j}^2+E_{J,j}\cos{\varphi_{q,j}}+C_{s}\dot{\Phi}_{s,j}^2+2E_{s}\cos{(\varphi^{(j)}_{ext})}\cos{\varphi_{s,j}}\bigg]+\bigg[\frac{C_R}{2}\dot{\Phi}_{C}^2-\frac{\Phi_{C}^2}{2L_{R}}\bigg]
\\\nonumber
&&+ \sum_{j=1}^2\left[\frac{C_c}{2}(\dot{\Phi}_{q,j}-\dot{\Phi}_{s,j})^2+\frac{C_c}{2}(\dot{\Phi}_{C}-\dot{\Phi}_{s,j})^2\right]\nonumber\\
=&&\frac{1}{2}\left[ \vec{\dot{\Phi}}^{\intercal} \hat{C} \vec{\dot{\Phi}} - \vec{\dot{\Phi}}^{\intercal} \hat{C}_g \vec{V}_g - \vec{V}_g^{\intercal} \hat{C}_g \vec{\dot{\Phi}}\right] - U(\vec{\Phi})
\label{Eq1}
\end{eqnarray}
where 
\begin{eqnarray}
&&\vec{\Phi}=
\begin{pmatrix}
\Phi_{q,1}\\ \Phi_{s,1} \\ \Phi_{c}\\ \Phi_{s,2}\\ \Phi_{q,2} 
\end{pmatrix},\quad\vec{V}_g=
\begin{pmatrix}
V_{g,1}\\ 0 \\ 0\\ 0\\ V_{g,2}
\end{pmatrix},\quad \hat{C}= 
\begin{pmatrix}
C_{qbt} & -C_c & 0 & 0 & 0 \\
-C_c & C_{sqd} & -C_c & 0 & 0 \\
0 & -C_c & C_{cav} & -C_c & 0 \\
0 & 0 & -C_c & C_{sqd} & -C_c \\
0 & 0 & 0 & -C_c & C_{qbt}
\end{pmatrix},\quad \hat{C}_g= 
\begin{pmatrix}
C_{g} & 0 & 0 & 0 & 0 \\
0 & 0 & 0& 0 & 0 \\
0 & 0 & 0& 0 & 0 \\
0 & 0 & 0& 0 & 0 \\
0 & 0 & 0& 0 & C_g
\end{pmatrix}\nonumber\\
&&U(\vec{\Phi})=\frac{\Phi_C^2}{2L_R}-\sum_{j=1}^2\left[E_{J,j}\cos{\varphi_{q,j}}+2E_{s}\cos{(\varphi^{(j)}_{ext})}\cos{\varphi_{s,j}}\right]
\end{eqnarray}
with $C_{qbt}=C_g+C_q+C_c$, $C_{sqd}=2(C_s+C_c)$, $C_{cav}=C_R+2C_c$, and $\varphi_{a,j}=2\pi\Phi_{a,j}/\Phi_0$, where $\Phi_0=h/2e$ is the superconducting flux quantum and $2e$ is the electrical charge of a Cooper pair. 

To calculate the Hamiltonian of our system, we start calculating the conjugate momenta $Q_k=\partial \mathcal{L}/\partial{\dot{\Phi_k}}$, and obtaining
\begin{eqnarray}
\vec{Q}=\hat{C}\vec{\dot{\Phi}} - \hat{C}_g\vec{V}_g
\end{eqnarray}
where
\begin{eqnarray}
&&\vec{Q}=
\begin{pmatrix}
Q_{q,1}\\ Q_{s,1} \\ Q_{c}\\ Q_{s,2}\\ Q_{q,2} 
\end{pmatrix}
\end{eqnarray}

Considering high plasma frequency approximation, we have that $\dot{\Phi}_{s,j}\ll\dot{\Phi}_{q,j},\dot{\Phi}_{C}$ and $\ddot{\Phi}_{s,j} ~\ll~ \ddot{\Phi}_{q,j},\ddot{\Phi}_{C}$, then the conjugate momenta takes the form
\begin{eqnarray}
Q_{q,j}&=&C_{qbt}\dot{\Phi}_{q,j} - Q_{g,j}\Rightarrow \dot{\Phi}_{q,j}=\frac{Q_{q,j}-Q_{g,j}}{C_{qbt}},\nonumber\\
Q_{C}&=&C_{cav}\dot{\Phi}_{C}\Rightarrow\dot{\Phi}_{C}=\frac{Q_C}{C_{cav}},\nonumber\\
Q_{s,j}&=& - C_c(\dot{\Phi}_{q,j} + \dot{\Phi}_{C})\Rightarrow Q_{s,j}=-C_c\left( \frac{Q_{q,j}-Q_{g,j}}{C_{qbt}} + \frac{Q_C}{C_{cav}} \right).
\label{approxQ}
\end{eqnarray}
Also, according to the Euler-Lagrange equations $\frac{\partial \mathcal{L}}{\partial \Phi_k} - \frac{d}{dt}\frac{\partial \mathcal{L}}{\partial \dot{\Phi}_k}=0$, we obtain
\begin{eqnarray}
\ddot{\Phi}_{q,j}&=&-\frac{2\pi E_{J,j}}{C_{qbt}\phi_0}\sin(\varphi_{q,j})\nonumber\\
\ddot{\Phi}_C&=&-\frac{1}{L_RC_{cav}}\Phi_C\nonumber\\
\sin(\varphi_{s,j})&=&\frac{C_c\phi_0}{4\pi E_s\cos(\varphi_{ext}^{(j)})}(\ddot{\Phi}_{q,j}+\ddot{\Phi}_C)
\label{approxPhi}
\end{eqnarray}
where we have used the high-plasma frequency approximation. Also, we can consider low-impedance approximation, which means that $\sin(\varphi_{s,j})\approx\varphi_{s,j}$, then
\begin{equation}
\varphi_{s,j}=-\frac{C_c\phi_0}{2\pi E_{s,j}^{eff}}\left[\frac{2\pi E_{J,j}}{C_{qbt}\phi_0}\sin(\varphi_{q,j})+\frac{1}{L_RC_{cav}}\Phi_C\right],
\end{equation}
where $E_{s,j}^{eff}=2 E_s\cos(\varphi_{ext}^{(j)})$. By applying the Legendre transformation $\mathcal{H}(\vec{\Phi},\vec{Q})= \vec{Q}^{\intercal}\hat{C}^{-1}(\vec{Q}+\hat{C}_g\vec{V}_g)-\mathcal{L}$, we obtain $\mathcal{H}(\vec{\Phi},\vec{Q}) = \frac{1}{2}(\vec{Q}+\hat{C}_g\vec{V}_g)^{\intercal}\hat{C}^{-1}(\vec{Q}+\hat{C}_g\vec{V}_g)+U(\vec{\Phi})$, so
\begin{eqnarray}
\mathcal{H}(\vec{\Phi},\vec{Q})=&&\sum_{j=1}^2\left[\frac{1}{2C_{qbt}}(Q_{q,j}-2e\bar{n}_{g,j})^2 - E_{J,j}\cos\varphi_{q,j}+\gamma_{q,j}\sin^2\varphi_{q,j}\right] + \frac{1}{2C_{cav}}Q_c^2 + \frac{\Phi_c^2}{2L_R} + \gamma_R\Phi_c^2\nonumber\\
&&-\sum_{j=1}^2\left[\bar{g}_j\sin\varphi_{q,j}\Phi_c\right] 
\end{eqnarray}
where we have used the approximations given by Eq.~(\ref{approxQ}) and Eq.~(\ref{approxPhi}), and we define
\begin{eqnarray}
\gamma_{q,j}=\frac{C_c^2 E_{J,j}^2}{2C_{qbt}^2E_{s,j}^{eff}},\quad \gamma_{R}=\frac{C_c^2\Phi_0^2}{8\pi^2C_{cav}^2L_R^2(E_{s,1}^{eff}+E_{s,2}^{eff})},\quad \bar{g}_j=\frac{C_c^2\Phi_0 E_{J,j}}{2\pi C_{cav}C_{qbt}L_RE_{s,j}^{eff}}
\end{eqnarray}
and also, we replace $C_{g}V_{g,j}=-2e\bar{n}_{g,j}$. We quantize the Hamiltonian by promoting the classical variables to quantum operators, $Q_{\alpha}\rightarrow\hat{Q}_{q,j}=2e\hat{n}_{q,j}$, $\varphi_{q,j}\rightarrow \hat{\varphi}_j$, with the commutation relation $[e^{i\hat{\varphi}_{q,j}},\hat{n}_{q,j}]=e^{i\hat{\varphi}_{q,j}}$. And for the resonator $Q_c\rightarrow\hat{Q}_c=\sqrt{\hbar\omega_c C_{cav}/2}(a^{\dagger}+a)$, $\Phi_c\rightarrow\hat{\Phi}_c=i\sqrt{\hbar\omega_cL_R/2}(a-a^{\dagger})$. Then our quantum Hamiltonian reads
\begin{equation}
H=\sum_{j=1}^2\left[4E_c(\hat{n}_{q,j}-\bar{n}_{g,j})^2 - E_{J,j}\cos\varphi_{q,j}+\gamma_{q,j}\sin^2\varphi_{q,j}\right] + \hbar\omega_ca^{\dagger}a-i\sum_{j=1}^2G_j\sin\hat{\varphi}_{q,j}(a-a^{\dagger})
\end{equation}
with $E_c=e^2/2C_{qbt}$ and $G_j=\bar{g}_j\sqrt{\frac{\hbar\omega_cL_R}{2}}$.

Using the charge number basis, \ie, $\hat{n}_{q,j}=\sum_{k}k\ketbra{k_j}{k_j}$ we have
\begin{equation}
\cos\hat{\varphi}_{q,j}=\frac{1}{2}\left(\sum_{k}\ketbra{k_j}{k_j+1} + \textrm{H.c}\right), \quad \sin\hat{\varphi}_{q,j}=-\frac{i}{2}\left(\sum_{k}\ketbra{k_j}{k_j+1} - \textrm{H.c}\right).
\end{equation}

Due to the anharmonicity of the system, we can use the two-level approximation to obtain
\begin{equation}
H=\sum_{j=1}^2\omega_{q_j}\sigma_j^z + \hbar\omega_ca^{\dagger}a-i\sum_{j=1}^2G_j\sigma_j^y(a-a^{\dagger}),
\label{S12}
\end{equation}
where $\omega_{q,j}\approx E_{J,j}$ \cite{Peng2021PhysRevLett}.
We use an external flux of the form $\varphi_{ext}^{(j)}=A_0^{(j)} + \varphi_{AC}^{(j)}(t)$, with $\varphi_{AC}^{(j)}(t)=A_1^{(j)}\cos(\mu_1^{(j)}t+\phi_1^{(j)})+ A_2^{(j)}\cos(\mu_2^{(j)}t+\phi_2^{(j)})$. If $|A_1^{(j)}|,~|A_2^{(j)}|\ll |A_0|$, then
\begin{eqnarray}
&&\frac{2}{E_{s,j}^{eff}}=\frac{1}{E_s\cos\varphi_{ext}^{(j)}}\approx \frac{1}{E_s\cos(A_0^{(j)})}\left[1+\frac{\sin(A_0^{(j)})}{\cos(A_0^{(j)})}\varphi_{AC}^{(j)}(t)\right]\nonumber\\
&&\Rightarrow G_j=G_0^{(j)}+G_1^{(j)}(t),\quad G_0^{(j)}=\frac{C_c^2\Phi_0E_{J,j}}{4\pi C_{cav}C_{qbt}E_{s}\cos (A_0^{(j)})}\sqrt{\frac{\hbar\omega_c}{2L_R}},\quad G_1^{(j)}(t)=G_0^{(j)}\frac{\sin(A_0^{(j)})}{\cos(A_0^{(j)})}\varphi_{AC}^{(j)}(t).
\end{eqnarray}

Now the Hamiltonian (\ref{S12}) in the interaction picture is given by
\begin{eqnarray}
H_I=\sum_{j=1}^2 G_0^{(j)}\left\{ \sigma_{j}^+a^{\dagger}e^{i\Delta_j t}\left[1+\frac{\sin(A_0^{(j)})}{\cos(A_0^{(j)})}\varphi_{AC}^{(j)}(t)\right]- \sigma_{j}^+ae^{i\delta_j t}\left[1+\frac{\sin(A_0^{(j)})}{\cos(A_0^{(j)})}\varphi_{AC}^{(j)}(t)\right]\right\} + \textrm{H.c}
\end{eqnarray}
with $\Delta_j=\omega_{q,j}+\omega_c$, and $\delta_j=\omega_{q,j}-\omega_c$. If the qubits and the cavity are off-resonant, then by adjusting the frequencies in $\varphi_{AC}(t)^{(j)}$, we can activate and deactivate the different terms. In particular, choosing $\mu_1^{(j)}=\Delta_j$ and $\mu_2^{(j)}=\delta_j$, we obtain
\begin{eqnarray}
H_I\approx \sum_{j=1}^2\bar{G}_0^{(j)}\left( \sigma_{j}^+a^{\dagger}e^{i\phi_1}- \sigma_{j}^+ae^{i\phi_2}\right) + \textrm{H.c}
\end{eqnarray}
after rotating wave approximation, where $\bar{G}_0^{(j)}=\frac{G_0^{(j)}\sin(A_0^{(j)})}{2\cos(A_0^{(j)})}$. Finally, fixing $\phi_1^{(j)}=0$ and $\phi_2^{(j)}=\pi$, we obtain
\begin{eqnarray}
H_I\approx \sum_{j=1}^2\bar{G}_0^{(j)}\sigma_j^x(a+a^{\dagger})
\end{eqnarray}
that correspond to the interaction term in the two-qubit Rabi model. We note that $\bar{G}_0^{(j)}$ depends on the constant component in the external flux, so it can be tuned independent of the qubit frequencies.
The circuit design allows us to activate and deactivate the rotating and counter-rotating terms independently. Then, we can generate the two-qubit Rabi model activating both terms at the same time, as well as the two-qubit JC model by the activation only of the rotating terms.

\section{Lindblad master equation for numerical simulation}
We use the following Lindblad form master equation
\begin{eqnarray}\label{lind}
\dot{\rho}=&&-i[H_{pq},\rho]+ \frac{\kappa}{2}(2a\rho a^\dag- a^\dag a\rho-\rho a_i^\dag a_i)+\sum_{j=1}^2 \frac{\gamma_j}{2}(2\sigma_j\rho \sigma_j^\dag- \sigma_j^\dag \sigma_j\rho-\rho \sigma_j^\dag \sigma_j)\nonumber\\&&+\sum_{j=1}^2 \frac{\gamma_{j\phi}}{2}(\sigma_{jz}\rho \sigma_{jz} - \rho)
\end{eqnarray}
to carry out the numerical simulation. Here, $\kappa$ is the photon decay rate of the $i$th resonator, consisting of the intrinsic part $\kappa_{in}$ and coupling part $\kappa_c$ with respect to the TL. $\gamma_j$  and $\gamma_{j\phi}$ are the energy relaxation rate and the dephasing rate of the $j$th qubit, respectively. We choose $\kappa_{in}=10^{-4}\omega$, $\gamma_j=10^{-5}\omega$,  $\gamma_{j\phi}=2\times10^{-5}\omega$.  Although the ultrastrong coupling regime is reached, we use this Lindblad from master
equation because the decoherences are extremely small there.  Meanwhile, the qubit singlet state, which is decoupled from the resonator, has been generated when $\kappa_{c}=10^{-1}\omega$ is turned on, and after the first photon is emitted, we will enter the JC coupling regime.

\section{Peculiarities of the special dark state $\vert \psi_R\rangle$}
There are two peculiarities of $|\psi_{R}\rangle$: 1. The energy level $\vert E\rangle$ very close to $|\psi_{R}\rangle$ at Jaynes-Cummings (JC) coupling regime in the spectrum (Fig. 1(a) in the main text) becomes degenerate with $|\psi_{R}\rangle$  when the rotating wave approximation is applied \cite{algebra}. 2. $\langle\psi_{E=\omega}|\dot{H}_R|\psi_{R}\rangle=0$, no matter how fast the parameters changes, which also holds when $\dot{H}_R$ replaced by $\dot{H}_{JC}$. In consequence, the adiabatic evolution along $\vert \psi\rangle$ is not restricted by the adiabatic theorem
\begin{equation}
\left|\frac{\langle E_{m}(t)|\dot{H}|E_{n}(t)\rangle}{(E_m-E_n)^2}\right|\ll1~~~
m\neq n ~~~t\in[0,T]
\end{equation} at the degeneracy point. Numerical results show the adiabatic evolution can be done in $68\omega^{-1}$ with fidelity $99\%$.

Here we prove $\langle\psi_{E=\omega}|\dot{H}_{R}|\psi_{R}\rangle=0$. In the adiabatic evolution of $H_{R}$ as shown in Fig. 2(b) in the main text with $\Delta_1+\Delta_2=\omega$, and $g_{1}=g_{2}=g$,
\begin{equation}\label{h2rd}
 \dot{H}_{R}=\dot{\Delta}_1\sigma_{1z}-\dot{\Delta}_1\sigma_{2z}+\dot{g}(a+a^\dag)\sigma_{1x} + \dot{g}(a+a^\dag)\sigma_{2x}  .
\end{equation}
So it is easy to find 
\begin{equation}\label{co}
\dot{H}_{R}|\psi_{R}\rangle=\frac{1}{{\cal
N}}(|\downarrow,\uparrow\rangle+
 |\uparrow,\downarrow\rangle)\left[(\dot{g}(\Delta_1-\Delta_2)-2\dot{\Delta}_1 g)|1\rangle
\right].
\end{equation}
Substituting $E=\omega$ into $(H_{R}-E)|\psi_{E=\omega}\rangle=0$ and projecting it into $\vert 1,\downarrow,\uparrow\rangle-\vert 1,\uparrow,\downarrow\rangle$, we have
\begin{eqnarray}
(\Delta_2-\Delta_1)\langle 1,\downarrow,\uparrow|\psi_{E=\omega}\rangle=(\Delta_1-\Delta_2)\langle 1\uparrow,\downarrow|\psi_{E=\omega}\rangle.
\end{eqnarray}
Therefore,
\begin{eqnarray}\label{coe}
 \langle 1,\downarrow,\uparrow|\psi_{E=\omega}\rangle=-\langle 1,\uparrow,\downarrow|\psi_{E=\omega}\rangle,
\end{eqnarray}
such that $\langle\psi_{E=\omega}|\dot{H}_{R}|\psi_{R}\rangle=0$ considering Eq. \eqref{co}, no matter how fast the parameters change. This result can be easily extended to the two-qubit JC model $H_{JC}$.

\section{One-photon special dark-state solution to the two-qubit quantum Rabi-Stark model}
We emphasize the deduction to obtain the one-photon dark-state solution to the two-qubit quantum Rabi-Stark model here. A similar deduction can be found in \cite{gaoxun}. The two-qubit Rabi-Stark model reads
\begin{eqnarray}\label{eq2}
H_{RS}=\omega a^{\dag }a + g_{1}{\sigma _{1x}}(a + {a^\dag }) + g_{2}{\sigma _{2x}}(a + a^{\dag }) + \Delta _{1}\sigma _{1z} + \Delta _{2}{\sigma _{2z}}+U_1 a^{\dag }a\sigma_{1z}+U_2 a^{\dag }a\sigma_{2z},
\end{eqnarray}
where $U_1$ and $U_2$ are couplings of the stark terms. The parity $P=\sigma_{1z}\sigma_{2z}\exp (i\pi a^\dag a)$ is still conserved here. Supposing there is a solution with at most one photon $\vert \psi_{RS}\rangle=c_0 \vert 0,\uparrow,\uparrow\rangle+c_1 \vert 0,\downarrow,\downarrow\rangle+c_2 \vert 1,\downarrow,\uparrow\rangle+c_3 \vert 1,\uparrow,\downarrow\rangle$ in the even parity subspace, then the eigenvalue equation $(H_{RS}-E)\vert \psi_{RS}\rangle=0$ reads
\begin{eqnarray}
\small\label{eigcq}\nonumber
  \left(\begin{array}{cccccc}
  \Delta_1+\Delta_2-E &0&g_{1} & g_{2} \\
   0&-\Delta_1-\Delta_2-E &g_{2}&g_{1} \\
   g_{1} & g_{2}& \omega-\Delta_1+\Delta_2-U_1+U_2-E&0\\
    g_{2} & g_{1}&0& \omega+\Delta_1-\Delta_2+U_1-U_2-E \\
           0&0&\sqrt{2}g_{1} & \sqrt{2}g_{2} \\
            0&0&\sqrt{2}g_{2} & \sqrt{2}g_{1} \\
  \end{array}\right) 
  \left(\begin{array}{c}
  c_0 \\
   c_1 \\
   c_2\\
   c_3\\
  \end{array}\right)=0 .
\end{eqnarray}
There are more equations than variables in this linear equation set, which normally does not have nontrivial solutions. However, quasi-exact solutions exist when parameters themselves satisfy certain relations. We apply elementary row transformations to the $6\times4$ matrix and find the nonzero rows can be less than the columns when $\Delta_1+\Delta_2=\omega=E$ and $g_1=g_2=g$, with the matrix transformed into
\begin{equation} \label{eigen2}
\begin{pmatrix}
   g &0 & -\Delta_1+\Delta_2-U_1+U_2&0\\
   0 & 1&0&0 \\
  0&0&1&1\\
   0&0 &0&0\\
    0&0&0&0\\
    0&0&0&0\\
\end{pmatrix},
\end{equation}
which means nontrival solution
\begin{equation}
\vert \psi_{RS}\rangle=(\Delta_1-\Delta_2+U_1-U_2)\vert 0\uparrow\uparrow\rangle+g\vert 1\rangle(\vert \downarrow\uparrow-\uparrow\downarrow\rangle)
\end{equation}
exist for arbitrary $g$ with constant energy, corresponding to a horizontal line in the spectrum, which is a special dark state. $\vert \psi_{RS}\rangle$ reduces to $\vert \psi_{R}\rangle$ when $U_1=U_2$.

\end{widetext}

\end{document}